# Intelligent Architectures for Intelligent Machines

Onur Mutlu
ETH Zurich
omutlu@gmail.com

## ABSTRACT

Computing is bottlenecked by data. Large amounts of application data overwhelm storage capability, communication capability, and computation capability of the modern machines we design today. As a result, many key applications' performance, efficiency and scalability are bottlenecked by data movement. In this keynote talk, we describe three major shortcomings of modern architectures in terms of 1) dealing with data, 2) taking advantage of the vast amounts of data, and 3) exploiting different semantic properties of application data. We argue that an intelligent architecture should be designed to handle data well. We show that handling data well requires designing architectures based on three key principles: 1) data-centric, 2) data-driven, 3) data-aware. We give several examples for how to exploit each of these principles to design a much more efficient and high performance computing system. We especially discuss recent research that aims to fundamentally reduce memory latency and energy, and practically enable computation close to data, with at least two promising novel directions: 1) performing massively-parallel bulk operations in memory by exploiting the analog operational properties of memory, with low-cost changes, 2) exploiting the logic layer in 3D-stacked memory technology in various ways to accelerate important data-intensive applications. We discuss how to enable adoption of such fundamentally more intelligent architectures, which we believe are key to efficiency, performance, and sustainability. We conclude with some guiding principles for future computing architecture and system designs.

## INTRODUCTION

Existing computing systems process increasingly large amounts of data. Data is key for many modern (and likely even more future) workloads and systems. Important workloads (e.g., machine learning, artificial intelligence, genome analysis, graph analytics, databases, video analytics), whether they execute on cloud servers or mobile systems are all data intensive; they require efficient processing of large amounts of data. Today, we can generate more data than we can process, as exemplified by the rapid increase in the data obtained in astronomy observations and genome sequencing [1].

Unfortunately, the way they are designed, modern computers are not efficient at dealing with large amounts of data: large amounts of application data greatly overwhelm the storage capability, the communication capability, and the computation capability of the modern machines we design today. As such, data becomes a large performance and energy bottleneck, and it greatly impacts system robustness and security as well. As a prime example, we provide evidence that the potential for new genome sequencing technologies, such as nanopore sequencing [2], is greatly limited by how fast and how efficiently we can process the huge amounts of genomic data the underlying technology can provide us with [3, 83].

The processor-centric design paradigm (and the ensuing processor-centric execution model) of modern computing systems is one prime cause of why data overwhelms modern machines [4, 5]. With this paradigm, there is a dichotomy between processing and memory/storage: data has to be brought from storage and memory units to compute units, which are far away from the memory/storage units. The dichotomy exists at the macro-scale (across the internet) and the micro-scale (within a single compute node). This processor-memory dichotomy leads to large amounts of data movement across the entire system, degrading performance and expending large amounts of energy. For example, a recent work [7] shows that more than 60% of the entire mobile system energy is spent on data movement across the memory hierarchy when executing four major commonly-used consumer workloads, including machine learning inference, video processing and playback, and web browsing. Similarly, due to the current design paradigm, a large fraction of the system resources is dedicated to units that store and move data, and actual computation units constitute only 5-20% of an entire chip [8] – yet, even then, data access is still a major bottleneck due to the large latency and energy costs of accessing large amounts of data.

## PRINCIPLES

Our starting axiom for an intelligent architecture is that it should handle (i.e., store, access, and process) data well. But, what does it mean for an architecture to handle data well? We posit (and later demonstrate with examples) that the answer lies in satisfying three major desirable properties (or principles): 1) data-centric, 2) data-driven, and 3) data-aware.

First, the system should ensure that data does not overwhelm its components. Doing so requires effort in intelligent algorithms, intelligent architectures and intelligent whole system designs that co-design across algorithms-architectures-devices, in a manner that puts data and its processing at the center of the design, minimizing data movement and maximizing the efficiency with which data is handled, i.e., stored, accessed, and processed (e.g., as exemplified in [4-38]). We call this first principle *data-centric architectures*.

Second, an intelligent architecture takes advantage of the vast amounts of data and metadata to continuously improve its decision making, by making both its policies and mechanisms better based on online learning. In other words, the architecture should make data-driven, self-optimizing decisions in its components (e.g., as exemplified in [39-51]). We call this second principle *data-driven architectures*.

Third, an intelligent architecture understands and exploits various properties of each piece of data so that it can improve and adapt its algorithms, mechanisms, and policies based on the characteristics of data. In other words, the architecture should make *data-characteristics-aware* decisions in its components and across the entire system (as exemplified in [52-58, 107, 116, 11]). We call this third principle *data-aware architectures*.

## COMPUTING ARCHITECTURES TODAY

Based on our qualitative and quantitative analyses, we find that existing computing architectures greatly fall short of handling data well. In particular they violate all of the three major desirable principles. We analyze each briefly next.

First, modern architectures are poor at dealing with data: they are designed to mainly store and move data, as opposed to actually compute on the data. Most system resources serve the processor (and accelerators) without being capable of processing data. As such, existing architectures are *processor-centric* as opposed to *data-centric*: they place the most value in the processor (not data) and everything else in the system is viewed as secondary serving the processor. We believe this is the wrong mindset and approach in designing a balanced system that handles data well: such a system should be data centric: i.e., data should be the prime thing that is valued and everything else in the system should be designed to 1) minimize data movement by enabling computation capability at and close to where data resides and 2) maximize the value and efficiency of processing data by enabling low-latency and low-energy access to as well as low-energy and low-cost storage of vast amounts of data.

Second, modern architectures are poor at taking advantage of vast amounts of data (and metadata) available to them during online operation and over time. They are designed to make simple decisions based on fixed policies, ignoring massive amounts of easily-available data. This is because existing architectural policies make *human-driven* decisions as opposed to *data-driven* decisions, and humans, by nature, do not seem capable of designing policies and heuristics that consider hundreds, if not thousands, of different variables that may be useful to examine to dynamically adapt online policies. It is instructive to notice that a memory controller, for example, keeps executing exactly the same fixed policy (e.g., FR-FCFS [59, 60], PAR-BS [61] or some other heuristic based policy [62-73, 117, 118]), during the *entire lifetime* of a system (for many many years!), regardless of the positive or negative impact of the resulting decisions on the system. The controller sees a vast amount of data even in the timeframe of a single millisecond (let alone years), yet it cannot learn from that data and adapt its policy because the policy is rigid and hardcoded by a human. This is clearly not intelligent: for example, as humans, we have the capability to learn from the past and adapt our actions accordingly to not repeat the same mistakes as in the past or to choose the best policy/actions that we believe will provide the highest benefits in the future. Enabling similar intelligence and far-sightedness in controller and system policies in an architecture is necessary for obtaining good performance and efficiency under a variety of system conditions and workloads.

Third, modern architectures are poor at knowing and exploiting different properties of application data. They are designed to treat all data as the same (except for a small set of specialized hints that provide some opportunity to optimize based on data characteristics in a limited manner that is very specific to the particular optimization). As such, the decisions existing architectures make are *component-aware* decisions as opposed to *data-aware* decisions: a component's (e.g., a cache's or a memory controller's) characteristics dominate the policies designed to control that component and the accessed data's characteristics are rarely conveyed to the policy or even known. If the characteristics of the data to be accessed or manipulated were known, the decisions taken could be very different: for example, if we knew the relative compressibility of different types of data [74-81], different components in the entire system could be designed in a manner that adaptively scales their capability to match the compressibility of different data elements, in order to maximize both performance and efficiency. Modifying the architecture and its interface to become richer and more expressive, and to include rich and accurate information on various properties of data that is to be processed, is therefore critical to customizing the architecture to the characteristics of the data and, thus, enabling intelligent adaptation of system policies to data characteristics.

## INTELLIGENT COMPUTING ARCHITECTURES

A major chunk of our talk describes in detail the characteristics of an intelligent computing architecture, by concrete examples and their empirical evaluation. This paper does not go into detail, but provides a brief overview with references to other works that exemplify such architectures. A detailed version of this talk can be found in [82].

### Data-Centric

A data-centric architecture has at least four major characteristics. First, it enables processing capability in or near where data resides (i.e., in or near memory structures), as described in detail in [4-6, 8, 38] and exemplified by [7-12, 14, 19, 20, 24, 27, 30, 34, 84, 108-113]. Second, it provides low-latency and low-energy access to data, as exemplified by [11-13, 15-18, 21, 23, 31-33, 84-86]. Third, it enables low-cost data storage and processing (i.e., high capacity memory at low cost, via techniques like new memory technologies, hybrid memory systems and/or compressed memory systems), as exemplified by [22, 87-96, 74, 76, 78, 107, 116]. Fourth, it provides mechanisms for intelligent data management (with intelligent controllers handling robustness, security, cost, etc.), as described in detail in [97-103, 116] and exemplified by, e.g., [104-106, 116].

### Data-Driven

A data-driven architecture enables the machine itself to learn the best policies for managing itself and executing programs. Controllers in such an architecture, when needed, are data-driven autonomous agents that automatically learn far-sighted policies. A prime example of such a controller is the *reinforcement learning based self-optimizing memory controllers* [39]. Such controllers can not only improve performance and efficiency under a wide variety of conditions and workloads but also reduce the hardware and system designer's burden in designing sophisticated controllers [39].

### Data-Aware

A data-aware architecture understands what it can do with and to each piece of data, and uses this information about data characteristics to maximize system efficiency and performance. In other words, it customizes itself (i.e., its policies and mechanisms) to the characteristics of the data it is dealing with. Such an architecture requires knowledge of various characteristics of different data elements and structures. Many semantic or other characteristics of data (e.g., compressibility, approximability, sparsity, criticality, access and security semantics, locality, latency vs. bandwidth sensitivity) are invisible or unknown to modern hardware and thus need to be communicated or discovered. We believe efficient and expressive software/hardware interfaces and mechanisms, as exemplified by X-Mem [52, 53] and the Virtual Block Interface [56] as well as other works [54, 55, 57, 58, 107, 116, 11], are promising approaches to creating general-purpose data-aware architectures.